 \documentclass[proof]{WileyASNA-v1}

\articletype{Article Type}%
\usepackage{pgfplots}
\usetikzlibrary{calc}
\pgfplotsset{compat=newest}

\colorlet{myblue}{blue!30}

\usepackage{tikz}
\usepackage{graphicx}
\usepackage{bigints}
\usepackage{pgfplots}
\usetikzlibrary{calc}
\pgfplotsset{compat=newest}

\colorlet{myblue}{blue!30}

\articletype{Article Type}%

\received{<day> <Month>, <year>}
\revised{<day> <Month>, <year>}
\accepted{<day> <Month>, <year>}

\begin{document}

\title{Branch-Cut Cosmology and the Bekenstein Criterion}%\protect\thanks{This is an example for title footnote.}}

\author[1]{Jos\'e A. de Freitas Pacheco}

\author[2,3]{C\'esar A. Zen Vasconcellos*}

\author[4,5]{Peter O. Hess}

\author[2]{Dimiter Hadjimichef}

\author[6]{Benno Bodmann}

\authormark{Jos\'e A. de Freitas Pacheco \textsc{et al}}

\address[1]{Observatoire de la C\^ote d'Azur, Nice, France}

\address[2]{Instituto de F\'isica, Universidade Federal do Rio Grande do Sul (UFRGS), Porto Alegre, Brazil}

\address[3]{International Center for Relativistic Astrophysics Network (ICRANet), Pescara, Italy}

\address[4]{Universidad Nacional Aut\'onoma de Mexico (UNAM), M\'exico City, M\'exico}

\address[5]{Frankfurt Institute for Advanced Studies (FIAS), Hessen, Germany}

\address[6]{Universidade Federal de Santa Maria (UFSM), Santa Maria, Brazil}

\corres{*C\'esar A. Zen Vasconcellos. \email{cesarzen@cesarzen.com}}

%\presentaddress{Present address}

\abstract[Abstract]{In this contribution we address the implications of the Bekenstein Criterion  in the branch-cut cosmology.
The impossibility of packaging energy and entropy according to the Bekenstein Criterion in a finite size makes the transition phase of the branch-cut cosmology very peculiar, imposing a topological leap between the contraction and expansion phases of the primordial universe  or a transition region similar to a wormhole, with space-time shaping itself topologically in the format of a helix-format around a branch point. Singularity means that there is no way for space-time to begin smoothly. The branch-cut cosmology alternatively proposes a non-temporal beginning at all, a pure space configuration, through a Wick rotation which replaces the imaginary time component by the temperature, the cosmological time. 
}


\maketitle

\section{Entropy in the Early Universe}

A challenging problem of the standard cosmological model and the standard model of particle physics is to explain the baryon asymmetry of the universe, $n^{A}_B$~\citep{Zyla2020}:
\begin{equation}
n^{A}_B = \frac{n_B}{S} = \frac{n^+_B - n^-_{\bar{B}}}{S} = (8.2 - 9.2) \times 10^{-11} \quad \mbox{at} \quad 95\% \quad C.L., 
\label{ba}
\end{equation}
with the
baryon density $n_B = n_B^+ - n_B^-$ representing the net value of baryons and anti-baryons and with $S$ denoting the total entropy density of the universe. The number of
baryons per unit of co-moving volume represents
a conserved quantity\footnote{This quantity is conserved if we disregard small variations in the entropy density when particles of the standard model annihilate, decreasing the effective number of degrees of freedom, heating photons (and other still coupled particles) but not particles that have already decoupled from the cosmic plasma. This occurs, for instance, when positrons and electrons annihilate. The released energy heats the photons but not the neutrinos, which have decoupled earlier.} since $n_B a^3(t) = constant$, where $a(t)$ is the cosmic scale factor.

In relativistic cosmology, the ``adiabatic expansion'' of the universe is related  to  
$s a^3(t) = constant$, where $s$ represents the entropy density; when the space-time singularity is approached, $a(t) \to 0$, the entropy density goes to ``infinity''.

The entropy inside the horizon is another quantity frequently computed, in spite of not satisfying the fundamental conservation laws of thermodynamics. Horizons are present in accelerated expanding universes (like for instance the de Sitter model). In static space-times, such as around black holes, the event horizon corresponds to a Killing surface, that is, a surface where a Killing vector field, $\xi_a$, associated to the
considered metric, is null. However, for most cosmological cases described by non-stationary space-times, Killing horizons do not exist! However, some investigations have been oriented towards local definitions of cosmological horizons~\citep{Hayward,Nielsen}; these alternative conceptions correspond to the apparent and the trapping horizons. The former is defined geometrically as a surface at which at least one pair of orthogonal null congruence of curves\footnote{In general relativity, a congruence of curves defines the set of integral curves of a vector field in a four-dimensional Lorentzian manifold which models spacetime.} have zero expansion~\citep{Bousso} whereas the latter corresponds to a compact space-like two-surface for which the expansion of one of the future-directed null normal vanishes~\citep{Nielsen2008}.  

Consider a flat Friedmann–Lemaître–Robertson–Walker (FLRW)~\citep{Friedmann1922, Lemaitre1927, Robertson1935, Walker1937}  spacetime whose metric is given by
\begin{equation}
    ds^2 = h_{mn} dx^m dx^n R^2 d \Omega^2 \,  , \label{metric}
\end{equation}
where $R = a(t)\, r$ is the radius of the two-sphere and $h_{mn} = diag[-1, a^2]$,
corresponding to the coordinates $x^0 = t$ and $x^1 = r$. The dynamical apparent horizon is defined
by the relation
\begin{equation}
h^{mn}\partial_m R \partial_nR = 0, \label{horizon}
\end{equation}
which implies that the vector $\xi_m = \partial_m R$ is null on the apparent horizon surface~\citep{Bak2000}.
 Using eqs. (\ref{metric}) and (\ref{horizon}), the apparent horizon radius reduces to $R_A = \frac{1}{H}$.
Hence, for a flat FLRW metric the apparent horizon coincides with the Hubble radius. During the radiative era, the apparent horizon varies with time as $R_ A = 2ct$. In this case, the entropy inside the apparent horizon is
\begin{equation}
S_A = \frac{2\pi^2 g_s}{45}\Biggl( \frac{kT}{\hbar c} \Biggr)^3 V_A.  \label{V}
\end{equation}
Since the temperature varies as $T \sim t^{-1/2}$ and the proper volume defined by the apparent radius
varies as $V \sim t^3$, one obtains from eq. (\ref{V}) that $S \sim t^{3/2}$. Consequently, the entropy goes to zero as 
the singularity is approached.

Suppose the existence of a minimum horizon radius of the order of the Planck length. In fact, Loop Quantum Cosmology (LQC) suppresses the big bang singularity (see \citep{Mercuri2010,Rovelli2010,Ashtekar2010}. In this theory, the universe is in a previous contraction phase, reaching a state of maximum (but finite) density (the “big bang”) and then expanding again. In the maximum contraction phase, the universe reaches a state of minimum area given by (see \citep{Mehdi}): 
\begin{equation}
A_{min} = 2 \pi \sqrt{3} \gamma \ell^2_P \simeq 2.61 \ell^2_P, 
\end{equation}
where $\gamma \simeq 0.23753$ is the so-called Immirzi-Barbero parameter. The equivalent horizon radius
corresponding to such a minimum surface is 
$R_{min} = (A_{min}/4 \pi)^{1/2} = 0.465 \ell_P$. Identifying this
radius with the apparent horizon at time $t_{min} = 0.228 t_P$, the
energy density of radiation at this instant can be now evaluated by using the Hubble equation,
that is
\begin{equation}
    \rho_{max} = \frac{3 c^2}{32 \pi G t^2_{min}} \simeq 0.574 \frac{\hbar c}{\ell^4_p}. \label{rhomax}
\end{equation}
It is interesting to compare this result with the energy density of a scalar field expected in LQC at this phase of maximum contraction, i.e.,
\begin{equation}
\rho_{LQC} = \frac{\sqrt{3}}{32 \pi^2 \gamma^3}\frac{\hbar c}{\ell^4_p}  \simeq 0.409 \frac{\hbar c}{\ell^4_p}.
\end{equation}
These values are quite comparable. The temperature at this instant can be computed from the equation
\begin{equation}
  \frac{\pi^2}{30} g_{eff} \frac{(kT)^4}{(\hbar c)^3} = \rho_{max}, 
\end{equation}
with $g_{eff} = 106.75$ for the standard model. One obtains $T_{max} = 5.09 \times 10^{31} K$ or $4.46 \times 10^{18}$ GeV (notice that this value is less than the Planck energy). Under these conditions, the dimensionless matter-energy entropy\footnote{Entropy can be defined to be dimensionless when temperature T is defined as an energy (dubbed tempergy).} inside the minimum volume is $0.850$ while the associated area entropy is $0.653$.

%%%%%%%%%%%%%%%%%%%%%%%%%%
\section{Bekenstein Bound}
%%%%%%%%%%%%%%%%%%%%%%%%%%

A key factor to understand the upper bound of entropy contained within a certain finite region of space with a finite amount of energy is the Bekenstein bound~\citep{Bekenstein1981}, a fundamental criterion which settles the basis for the generalization of entropy and the second law of thermodynamics for non-gravitational systems. Applied to the primordial universe by considering the connected spatial region within
the particle horizon of a given observer, i.e., the locus
of the most distant points that can be observed at a specific time $t_0$ in an event, Bekenstein
conjectured an upper bound, given by  $\frac{2 \pi R}{\hbar c}$, for  the entropy $S$ and energy $E$ of a system enclosed in a spherical region of radius $R$:  
\begin{eqnarray}
    \frac{2 \pi R}{\hbar c} \geq S/E \quad \mbox{so} \quad S \leq S_B = \frac{2 \pi}{\hbar c} E R,  \label{S}
\end{eqnarray}
with $S_B$ denoting the upper limit of the Bekenstein bound.  

The implications of the Bekenstein Criterion are striking\footnote{According to the Bekenstein limit, for a self-gravitating body, the state of maximum entropy corresponds to a black hole!}, establishing initial physical conditions that impact the evolution, symmetries and conservation laws of elementary particles in the early universe. The fulfillment of the criterion would imply  a non-singular, isotropic and homogeneous primordial universe 
with entropy, temperature, and baryon number equal to zero. Moreover, in a scenario in which the baryon asymmetry of the universe is induced by an asymmetry of leptons generated in the decays of heavy sterile neutrinos (leptogenesis), the primordial lepton number must also be equal to zero. Evidently, in a quantum treatment, based on the premise of a universe that `materializes out of nothing' as a result of quantum fluctuations, the extent of the particle horizon will never be smaller than the minimum size dictated by the radius of curvature of the emerging spacetime, i.e., of the order of the Planck length~\citep{Powell2020}, at the scale where general relativity and quantum mechanics are expected to merge.  This could represent a mechanism to suppress the presence of singularities in the early universe.

In order to check if the Bekenstein upper limit is violated or not, 
notice that during the radioactive era, the Bekenstein entropy limit varies as $S \sim t^2$ meaning that it goes 
also to zero as the singularity approaches. On the other side, since the matter-radiation entropy
$ S_{m+r} \sim t^{3/2}$ (as previously seen), one should expect that close to the singularity the limit will be violated and only after a critical instant $t_c$ it will be
satisfied. Let us verify if at the adopted minimum time the limit is violated or not. The
energy inside the horizon at this instant is $E = \rho_{max} \bigl( \frac{4 \pi}{3} \bigr)R^3_{min}$.
Replace this into eq. (\ref{S}) and use
eq. (\ref{rhomax}) to obtain
\begin{equation}
    S_{B_*} = \frac{8 \pi^2}{3} (0.574)(0.456)^4 = 0.653.
\end{equation}
Hence, under these conditions, the Bekenstein limit is violated since the mater-radiation entropy
is $0.850$ but not for the area entropy since $S_A = S_B$ . The Bekenstein limit and the area entropy
varies as the square of time, hence the equality will be always satisfied during the radiation era. At which instant the limit will the obeyed? Using the time dependence of both quantities one obtains
\begin{equation}
    0.850 \, \Bigl(\frac{t}{t_m} \Bigr)^{3/2} =  0.653 \, \Bigl(\frac{t}{t_m} \Bigr)^2 \Rightarrow t = 1.694 \, t_{min} =  0.386 \, t_P. 
\end{equation}
Hence, just after $t_{min}$ the Bekenstein limit is valid again.

In the following we address the implications of the Bekenstein Criterion and the mutual consistency of the standard cosmological model, the standard model of particle physics, and General Relativity in the branch cut cosmology.

%%%%%%%%%%%%%%%%%%%%%%%%%%%%%%%
\section{Branch Cut Cosmology: Bekenstein Criterion and entropy} 
%%%%%%%%%%%%%%%%%%%%%%%%%%%%%%%

In the standard cosmology, the patch corresponding to the observable universe was never causally connected in the past~\citep{Ijjas2014}. 
In the present time ($t = t_0$), the patch size, $R(t_0)$ and the horizon size, $H^{-1}(t_0)$, are equal, ie, $R( t_0) = H^{-1}(t_0)$. In earlier times, the ratio between the horizon size to the patch size decreases monotonically extrapolating back in time as $a(t) \rightarrow 0$ in the form 
$a^{\epsilon}(t)/a(t)$:  
\begin{equation}
\frac{H^{-1}(t)}{a(t)} \sim 
\frac{a^{\epsilon}(t)}{a(t)} = a^{\epsilon - 1}(t) \quad
\mbox{and} \quad  \lim_{a(t) \rightarrow 0}{a(t)^{\epsilon - 1}} \rightarrow 0, \label{aepsilona}
\end{equation}
with the horizon size approaching zero faster than the patch size.
In this equation $\epsilon(t)$ represents the dimensionless {\it thermodynamics connection} between the energy density $\rho(t)$ and the pressure $p(t)$ of a perfect fluid thus enabling the fully description of the equation of state (EoS) of the system
\begin{equation}
 \epsilon(t) \equiv \frac{3}{2} \Biggl(1 + \frac{p(t)}{c^2 \rho(t)}  \Biggr). \label{r0}
\end{equation}
According to the CMB measurements, the density and temperature were almost uniform throughout the primordial patch (last CMB surface scattering). Explaining the uniformity of the CMB at length scales greater than the size of the horizon at the last scattering surface 
and at all previous times constitutes the horizon problem~\citep{Ijjas2019}. The CMB measurements also reveal a spectrum of small amplitude density fluctuations, nearly scale-invariant whose explanation constitutes the inhomogeneity problem~\citep{Ijjas2014}.
Combined with the primordial singularity, where any trace of causality completely disappears, these two factors represent the main roots for solving the cosmic singularity, horizon, inhomogeneity and flatness problems of standard cosmology~\citep{Zen2022}. 

In dealing with these problems, the branch cut cosmology offer an alternative to overcome the presence of a primordial singularity and for the inflationary model of the universe and in addition propose a solution for the non-causal behavior of patch size and horizon size in standard cosmology~\citep{Zen2020,Zen2021a,Zen2021b}. 

In the branch cut cosmology, by means of the combination of the multiverse proposal by~\citet{Hawking2018} of a hypothetical set of multiple universes, existing in parallel and the technique of analytical continuation in complex analysis applied to the FLRW metric,  the Friedmann's field equations for a version of the  $\Lambda$CDM  ($\Lambda\neq 0$) model extended to the complex domain leads to a  new complex cosmic factor: 
%\begin{equation}
 $   \ln^{-1}[\beta(t)]$ with $ \beta(t) \simeq \frac{a(t) - \chi(t)}{a(t) + \chi(t)},$
%\end{equation}
and with the function $\chi(t)$ characterizing the range of $\ln^{-1}[\beta(t)]$ associated to the cuts in the branch cut. More precisely, $\ln^{-1}[\beta(t)]$ denote solutions of Einstein's equations for a FLRW-type metric extended to the complex plane, represented as the reciprocal
of a complex multi-valued function, the natural complex logarithm function $\ln[\beta(t)]$, not its inverse. 
The $\ln[\beta(t)]$ function corresponds to a helix-like superposition of cut-planes, the Riemann sheets,
with an upper edge cut in the n-th plane joined with a lower edge of cut in the (n + 1)-th plane, mapping Riemann sheets onto horizontal strips, which represent in the
branch-cut cosmology the time evolution of the time-dependent horizon sizes. The patch sizes
in turn maps progressively the various branches of the $\ln[\beta(t)]$ function which are glued along
the copies of each upper-half plane with their copies on the corresponding lower-half planes. In
the branch-cut cosmology, the cosmic singularity is replaced by a family of Riemann sheets in
which the scale factor shrinks to a finite critical size, — the range of $\ln[\beta(t)]$, associated to
the cuts in the branch cut, shaped by the $\chi(t)$ function —, well above the Planck length.  In this configuration, a(t) represents a component of a generalized new form factor associated with the combination of all multiverses confined to a single universe.

In short, in addition to branch cuts, there are `singularities', --- the branch points ---, but at the same time there are multiple points that configure continuous paths in the Riemann sheets. This enables continuous solutions off the primordial singularity, which, in general relativity, are inescapable, under the presumption at the level of a local continuity prevails, i.e.,  that there is some neighborhood of the branch point, let's call it $z_0$, close enough although not equal to $z_0$, where one can find a small local patch where $\ln^{-1}[\beta(t)]$ is single valued and continuous.  The largest region possible for the range is crucial to model the generation of the structures observed today via primordial fluctuations. 

Adopting the Bekenstein's criterion,
the largest region possible for this range, $\chi(t)$,  obeys
\begin{equation}
    \frac{2 \pi \chi(t) {\cal D}_0}{\hbar c} \geq S/E, \quad \Longrightarrow \quad \chi(t) \geq \frac{\hbar cS}{2 \pi {\cal D}_0 E},
    \label{Bekenstein}
\end{equation}
where ${\cal D}_0(t)$ is the proper distance (or proper length) at a reference time $t_{0}$. 
Assuming the Bekenstein Criterion, \begin{equation} \ln^{-1}[\beta(t)] \to
\ln^{-1}\Biggl[\frac{a(t) + \frac{\hbar cS}{4 \pi {\cal D}_0 E}}{a(t) - \frac{\hbar cS}{4 \pi {\cal D}_0 E}}\Biggr] \label{Bekenstein2}.
\end{equation}
Evidently the complex nature of the cosmic factor $\ln[\beta(t)]$ 
extends to all other components of the Bekenstein Criterion. In order to have a numerical 
assessment of the implications of the criterion, we consider in the following the modulus of the enumerated complex quantities.
According to the second law of thermodynamics, the entropy of the universe is always increasing, raising the possibility that, at the beginning of the universe, its value would be minimal. Zero entropy does not conform to Bekenstein Criterion. 

Next, we carry out a study of the implications of the Bekenstein criterion in the conformation of the thermodynamic parameters of the early universe. To this end, we consider two cases described below.

In the first case,
we assume an apparent horizon radius at the instant $t_*$ equal to the Planck length, that is,
$R_A = 2ct_* = \ell_P$. This means that $t_* = t_P/2$. Notice that one assumes a flat FRW model and that
the universe in these early phases is radiation-dominated by Standard Model particles, corresponding to an effective number of degrees of freedom $= 106.75$. Since the scale factor varies as $\ln[\beta(t_*)] \sim t^{1/2}$ (see \citep{Zen2021a}, the Hubble parameter varies as $H \sim 1/2t$. Hence, the Hubble equation at the instant t* can be written as
\begin{equation}
    H^2 =   \frac{1}{4t^2_*} =  \frac{1}{t^2} = \frac{8 \pi G}{3 c^2} \rho_*. \label{rho*}
\end{equation}
From this equation, the energy density can be estimated to be $\rho_* = 5.55 \times 10^{111} erg/cm^3$.
The temperature can be now calculated from the thermodynamic relation for the energy density
of relativistic particles
\begin{equation}
    \rho_* = \frac{\pi^2}{30} g_{eff} \frac{\bigl(kT_*\bigr)^4}{\bigl(\hbar c\bigr)^3} \Rightarrow T_* = 3.43
    \times 10^{31} K (= 2.95 \times 10^{18} GeV). 
\end{equation}
The radiation energy inside the apparent horizon is
\begin{equation}
    E_* = \frac{4 \pi}{3} \ell^3_P
\rho_* = 2.32 \times 10^{13} erg (=1.45 \times 10^{25} eV).
\end{equation}
The entropy of radiation inside the apparent horizon is
\begin{equation}
    S_* = \frac{2 \pi^2}{45} g_{eff} \Bigl(\frac{kT_*}{\hbar c}\Bigr)^3 \times \frac{4 \pi}{3} \ell^3_P = 2.76.
\end{equation}
Now the Bekenstein limit can be calculated as follows – first, notice that eq. (\ref{rho*}), after some
algebra, can be written as
\begin{equation}
    \rho_* = \frac{3}{8 \pi}\frac{\hbar c}{\ell^4_P}.
\end{equation}
The Bekenstein limit is
\begin{equation}
    S_B(t_*) = \frac{2 \pi}{\hbar c}E_* \ell_P = \frac{2 \pi}{\hbar c} 
    \Bigl(\rho^* \frac{4 \pi}{3} \ell^3_P \Bigr) \ell_P = \pi.  
\end{equation}

%%%%%%%%%%%%%%%%%%%%TABLE%%%%%%%%%%%%%%%%%%%%%%%%%%%%%%%%%%%%%%%%%%%%%%%%%%%%
\begin{table}[htbp]
    \centering
    \begin{tabular}{|c|c|}
    \hline 
      Parameter & Value (SI units)   \\ \hline \hline
          $|T_U| = T_*$ &  $3.43 \times 10^{31} K$ \\ \hline
          $|E_U| = E_* $ &  $1.45 \times 10^{25} \, eV$ \\ \hline
            $|S_U| = S_*$ & $2.76$ \\ \hline
             $|\chi||{\cal D}_0|$ &  $\geq 5.98 \times 10^{-18}$ fm \\ \hline
    \end{tabular}
    \caption{Parameter values corresponding to the first case.}
    \label{table0}
\end{table}
%%%%%%%%%%%%%%%%%%%%%%%%%%%%%%%%%%%%%%%%%%%%%%%%%%%%%%%%%%%%%%%%%%%%%%%%%%%%%%

In the second case, the thermodynamic parameters when $z = 6.34$ will be now evaluated. The temperature is given by $T = 2.725 (1+z)$, which
indicates the adiabatic expansion of the universe. Notice that for red-shift smaller than $3300$, the universe is matter-dominated and the apparent horizon should be evaluated differently.
The Hubble parameter is now given by
\begin{equation}
    H(z) = H_0 \sqrt{\Omega_{\Lambda} + \Omega_m(1+z)^3}.
\end{equation}
Using Planck Collaboration parameters~\citep{Planck2018} (see Table \ref{2}) 
one obtains for the Hubble parameter at $z = 6.34$, $H=759.8 km/s/Mpc$, corresponding to an apparent horizon radius $R_A = 1.22 \times 10^{27} cm$.
\begin{table}[]
    \centering
   \begin{tabular}{|c|c|}
    \hline 
      Parameter & Value (SI units)   \\ \hline \hline
     $H_0$    & $68 km/s/Mpc$ \\  \hline
      $\Omega_{\Lambda}$   & $0.686$ \\  \hline
       $\Omega_m$ &  $0.314$ \\  \hline
    \end{tabular}
    \caption{Planck Collaboration parameters~\citep{Planck2018}}
    \label{2}
\end{table}
The matter energy density can be computed as before using the Hubble equation, that is
\begin{equation}
\rho_m = \frac{3 c^2}{8 \pi G}H^2(z) = 9.74 \times 10^{-7} erg/cm^3. \label{med}
\end{equation}
The energy is obtained simply by multiplying eq. (\ref{med}) by the volume defined by the horizon, that is,
$E_m = 7.41 \times 10^{75} erg (= 4.63 \times 10^{87} eV$). Then, the Bekenstein limit results to be
\begin{equation}
    S_B(z) = \frac{2 \pi}{\hbar c} \Biggl(\rho_m \frac{4 \pi}{3} R^3_A \Biggr) R_A = 1.8 \times 10^{120}.
\end{equation}
The internal entropy  due to matter is several orders of magnitude smaller than the entropy of
relativistic matter constituted only by photons and neutrinos (degrees of freedom $g_S =3.91$). Consequently the internal entropy is
\begin{equation}
    S_{rad} = \frac{2 \pi^2}{45} g_S \Bigl(\frac{kT}{\hbar c} \Bigr)^3 \frac{4 \pi}{3} R^3_A
    = 8.49 \times 10^{87}.
 \end{equation}

%%%%%%%%%%%%%%%%%%%%%%%%%%%%%%%TABLE%%%%%%%%%%%%%%%%%%%%%%%%%%%%%%%%%%%%%%%%%%%%%%%%%
\begin{table}[htbp]
    \centering
    \begin{tabular}{|c|c|}
    \hline 
      Parameter & Value (SI units)   \\ \hline \hline
          $|E_U| = E_m$  &  $  4.63 \times 10^{87}$ eV \\ \hline
                     $|S_U| = S_{rad}$ & $8.49 \times 10^{87}$   \\ \hline
                        $|\chi||{\cal D}_0|$ & $\geq 5.80 \times 10^{7}$ fm \\ \hline
    \end{tabular}
    \caption{Parameter values corresponding to the second case.}
    \label{table}
\end{table}
%%%%%%%%%%%%%%%%%%%%%%%%%%%%%%%%%%%%%%%%%%%%%%%%%%%%%%%%%%%%%%%%%%%%%%%%%%%%%%%%%%%%%%
The results obtained indicate, in the both cases considered, higher values than the Planck scale for  $|\chi||{\cal D}_0|$, reinforcing the conception of a topological leap in the transition region between the contraction and expansion phases in the first scenario of the branch cut cosmology.
A more accurate calculation would involve deeper knowledge about unstable dark matter, primordial black holes, strong primordial heterogeneities, which may play an important role in identifying the primordial energy.
%%%%%%%%%%%%%%%%%%%%%%%%%%%%%%%%%%%%%%FIGURE%%%%%%%%%%%%%%%%%%%%%%%%%%%%%%%%%%%%%%%
\begin{figure*}[htbp]
\centering
\includegraphics[width=80mm,height=50mm]{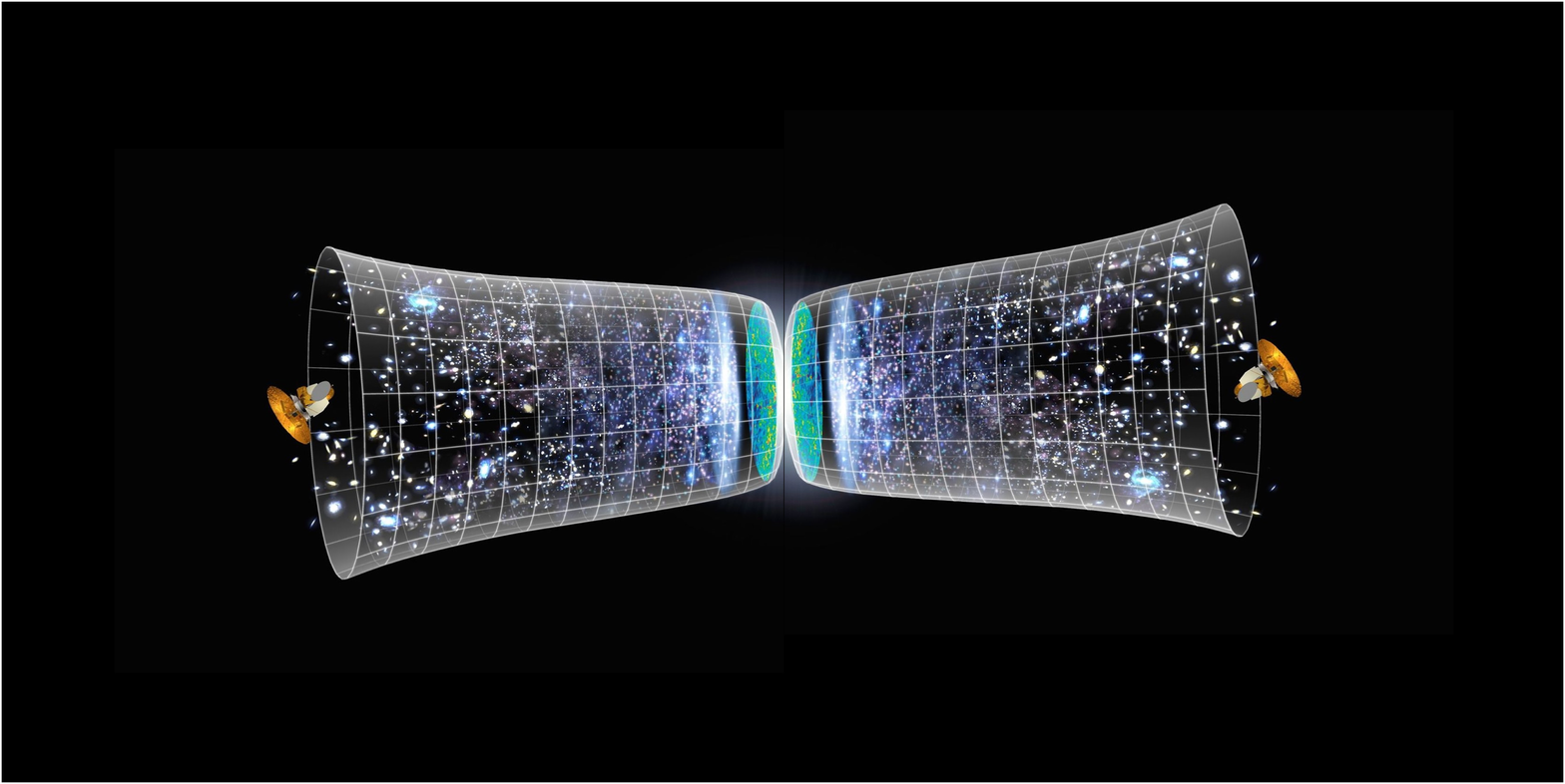}   \label{ESO1} \hspace{0.75cm}
\includegraphics[width=80mm,height=50mm]{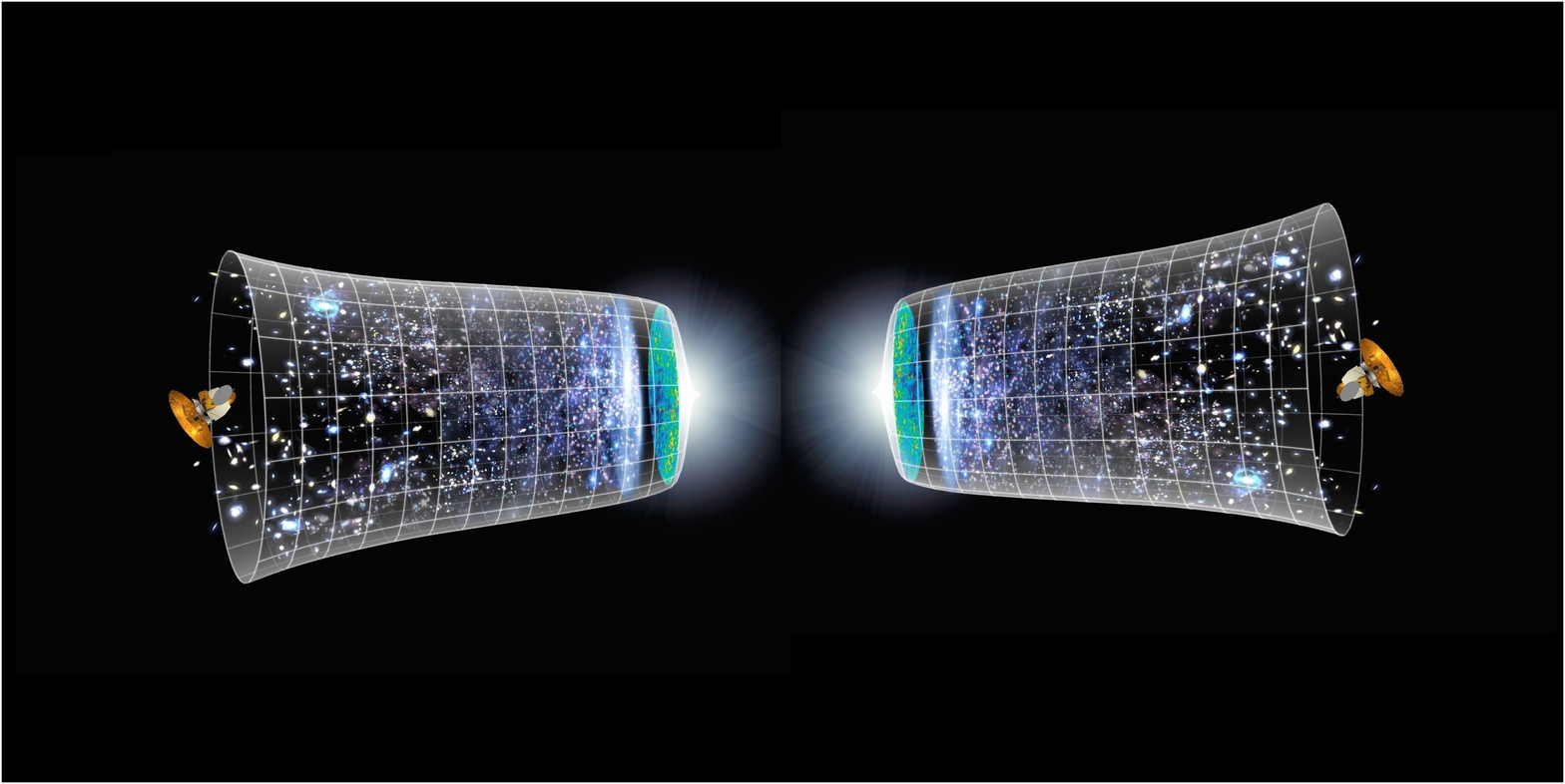}  \label{ESO2}
\caption{The figure shows artistic representations of the scenarios in the branch-cut cosmology. The first scenario, sketches the cosmic contraction and expansion phases of the branch cut universe evolution with no primordial singularity. In the second scenario there are two primordial singularities and a mirror universe nested to ours. Images created on basis of the original images from~\citet{Kormesser2020}\label{ESO}.}
\end{figure*}
%%%%%%%%%%%%%%%%%%%%%%%%%%%%%%%%%%%%%%%%%%%%%%%%%%%%%%%%%%%%%%%%%%%%%%%%%%%%%%%%%%%%

%%%%%%%%%%%%%%%%%%%%%%%%%%%%%%%%%%%%%%%%%%%%%%%
\section{Scenarios of the branch-cut cosmology}
%%%%%%%%%%%%%%%%%%%%%%%%%%%%%%%%%%%%%%%%%%%%%%%

The results presented so far in Tables \ref{table0} and \ref{table} are consistent with the idea conceived by~\citet{Bekenstein1981,Bekenstein2003}, originally applied to black holes and later extended to the primordial universe, demonstrating that it is physically impossible to 'pack' large values of entropy in a region with a certain limited area, or in a certain mass with a defined extension. These conclusions impact our evolutionary view of the universe and point to a few important conclusions.  

In information theory, the so called information equation, defined as
\begin{eqnarray}
I(p) = - log_b(p)\, ,    
\end{eqnarray}
relates the degree of information associated to a particular event, $I(p)$, and 
the probability $p$ that this event may occur, with the values of the dimensionless entropy limited to the range $0 \leq entropy \leq log(n)$, where $n$ represents the number of outcomes. Minimum entropy corresponds to maximum probability of a certain event to occur; conversely, maximum entropy occurs when all probabilities of all outcomes have equal values, more precisely, $1/n$. This conception reinforces the idea that the entropy at the beginning of the universe is close to zero, and cannot be null since, according to the Bekenstein Criterion, singularities would be, from a thermodynamic point of view, impossible to occur. Recalling that the Bekenstein Criterion imposes that the initial state of the universe is unique, therefore, in a probabilistic conception, the primordial state of the universe would fit the case of minimum entropy as theorized in the theory of information. In the process of formation of a black hole, the catalyzed conversion of a pure quantum state to a mixed state occurs, in contradiction with the principle of unitary quantum evolution, thus causing loss of information. This realization has led to the `information paradox', a topic that has been the scene of fierce conceptual disputes.
The results from Tables \ref{table0} and \ref{table} indicate that even considering much lower values of temperature, thermal energy and entropy than the values adopted on the Planck scale, the `extension' of the branch point in the branch cut cosmology should be about $10^{30}$ times greater than the corresponding value in the Planck scale.

Two scenarios have been delineated for the evolution of the branch cut cosmology which are
sketched in an artistic representation 
(see Fig. \ref{ESO}), with a branch point and a branch cut on the left figure, and two primordial singularities on the right figure\footnote{Figures based on an artistic impression originally developed by ESO / M. Kornmesser~\cite{Kormesser2020}.}. 
In the first scenario of the branch-cut cosmology, the universe evolves continuously from
the negative complex cosmological time sector to the positive
one, circumventing continuously a branch cut in the transition region, and no primordial singularity occurs in the
imaginary sector, only branch points. 
In the second scenario, the branch cut and branch point disappear after the {\it realisation} of imaginary time by means of a Wick rotation, which is replaced here by the real and continuous thermal time (temperature). In this second scenario, a mirrored parallel evolutionary universe, adjacent to ours, is nested in the structure of space and time, with its evolutionary process going backwards in the cosmological thermal time negative sector. In this case, the connection between the previous solutions is {\it broken} as a result of the Wick rotation.  In the contraction phase (first scenario), as the patch size decreases with a linear dependence on $\ln[\beta(t)]$,
light travels through geodesics on each Riemann sheet, circumventing continuously the branch 
cut, and although the horizon size scales with $\ln^{\epsilon}[\beta(t)]$ and the patch size in turn scales as $\ln[\beta(t)]$, the length of the path to be traveled by light circumventing the branch cut compensates for the scaling difference between the patch and horizon sizes. Under these circumstances, causality between the
horizon size and the patch size may be achieved through the accumulation of branches in the
transition region between the present state of the universe and the past events. This topic, the number of branches accumulated in order to reach causality, is a topic that deserves our attention in a following investigations.

Another topic that deserves attention is the main conclusion of this contribution: the impossibility of packaging energy and entropy according to the Bekenstein Criterion in a finite size makes the transition phase very peculiar, imposing a topological leap between the two phases or a transition region similar to a wormhole, with space-time shaping itself topologically in the format of a helix-shape like as proposed by branch-cut cosmology around a branch-point. 

Singularity means that there is no way for space-time to begin smoothly. The branch-cut cosmology alternatively proposes a non-temporal beginning at all, a pure space configuration, through a Wick rotation which replaces the imaginary time component by the temperature, the cosmological time. 

\section*{Acknowledgements}

P.O.H. acknowledges financial support from PAPIIT-DGAPA (IN100421).

%%%%%%%%%%%%%%%%%%%%%%%%%%%%%%%%%%%%%%%%%%%%%%%%%%%%%
%%%%%%%%%%%%%%%%%%%%%%%%%%%%%%%%%%%%

%\bibliographystyle{wileyNJD-VANCOUVER}

\bibliography{Zen.bib}%

\begin{thebibliography}{}

\bibitem [\protect \citeauthoryear {%
Aghanim%
\ \BBA {} et al.%
}{%
Aghanim%
\ \BBA {} et al.%
}{%
{\protect \APACyear {2020}}%
}]{%
Planck2018}
\APACinsertmetastar {%
Planck2018}%
\begin{APACrefauthors}%
Aghanim, N.%
\BCBT {}\ \BBA {} et al.%
\end{APACrefauthors}%
\unskip\
\newblock
\APACrefYearMonthDay{2020}{}{},
\newblock
\unskip
\newblock
\APACjournalVolNumPages{Astronomy \& Astrophysics}{641}{}{A6}.
\PrintBackRefs{\CurrentBib}

\bibitem [\protect \citeauthoryear {%
Ashtekar%
\ \BBA {} Sloan%
}{%
Ashtekar%
\ \BBA {} Sloan%
}{%
{\protect \APACyear {2010}}%
}]{%
Ashtekar2010}
\APACinsertmetastar {%
Ashtekar2010}%
\begin{APACrefauthors}%
Ashtekar, A.%
\BCBT {}\ \BBA {} Sloan, D.%
\end{APACrefauthors}%
\unskip\
\newblock
\APACrefYearMonthDay{2010}{}{},
\newblock
\unskip
\newblock
\APACjournalVolNumPages{Phys. Lett. B}{694}{}{108}.
\PrintBackRefs{\CurrentBib}

\bibitem [\protect \citeauthoryear {%
Assanioussi%
, Dapor%
, Liegener%
\BCBL {}\ \BBA {} Pawlowski%
}{%
Assanioussi%
\ \protect \BOthers {.}}{%
{\protect \APACyear {2019}}%
}]{%
Mehdi}
\APACinsertmetastar {%
Mehdi}%
\begin{APACrefauthors}%
Assanioussi, M.%
, Dapor, A.%
, Liegener, K.%
\BCBL {}\ \BBA {} Pawlowski, T.%
\end{APACrefauthors}%
\unskip\
\newblock
\APACrefYearMonthDay{2019}{}{},
\newblock
\unskip
\newblock
\APACjournalVolNumPages{Phys. Rev. D}{100}{}{084003}.
\PrintBackRefs{\CurrentBib}

\bibitem [\protect \citeauthoryear {%
Bekenstein%
}{%
Bekenstein%
}{%
{\protect \APACyear {1981}}%
}]{%
Bekenstein1981}
\APACinsertmetastar {%
Bekenstein1981}%
\begin{APACrefauthors}%
Bekenstein, J.%
\end{APACrefauthors}%
\unskip\
\newblock
\APACrefYearMonthDay{1981}{}{},
\newblock
\unskip
\newblock
\APACjournalVolNumPages{Int. J. Theor. Phys.}{28}{}{967–989}.
\PrintBackRefs{\CurrentBib}

\bibitem [\protect \citeauthoryear {%
Bekenstein%
}{%
Bekenstein%
}{%
{\protect \APACyear {2003}}%
}]{%
Bekenstein2003}
\APACinsertmetastar {%
Bekenstein2003}%
\begin{APACrefauthors}%
Bekenstein, J.%
\end{APACrefauthors}%
\unskip\
\newblock
\APACrefYearMonthDay{2003}{}{},
\newblock
\unskip
\newblock
\APACjournalVolNumPages{Contemp.Phys.}{45}{}{31-43}.
\PrintBackRefs{\CurrentBib}

\bibitem [\protect \citeauthoryear {%
Bousso%
}{%
Bousso%
}{%
{\protect \APACyear {2002}}%
}]{%
Bousso}
\APACinsertmetastar {%
Bousso}%
\begin{APACrefauthors}%
Bousso, R.%
\end{APACrefauthors}%
\unskip\
\newblock
\APACrefYearMonthDay{2002}{}{},
\newblock
\unskip
\newblock
\APACjournalVolNumPages{Phys. Rev. D}{74}{}{825}.
\PrintBackRefs{\CurrentBib}

\bibitem [\protect \citeauthoryear {%
D.Bak%
\ \BBA {} S.J.Rey%
}{%
D.Bak%
\ \BBA {} S.J.Rey%
}{%
{\protect \APACyear {2000}}%
}]{%
Bak2000}
\APACinsertmetastar {%
Bak2000}%
\begin{APACrefauthors}%
D.Bak%
\BCBT {}\ \BBA {} S.J.Rey.%
\end{APACrefauthors}%
\unskip\
\newblock
\APACrefYearMonthDay{2000}{}{},
\newblock
\unskip
\newblock
\APACjournalVolNumPages{Class. Quant. Grav.}{17}{}{L3}.
\PrintBackRefs{\CurrentBib}

\bibitem [\protect \citeauthoryear {%
Ding%
\ \BBA {} Rovelli%
}{%
Ding%
\ \BBA {} Rovelli%
}{%
{\protect \APACyear {2010}}%
}]{%
Rovelli2010}
\APACinsertmetastar {%
Rovelli2010}%
\begin{APACrefauthors}%
Ding, Y.%
\BCBT {}\ \BBA {} Rovelli, C.%
\end{APACrefauthors}%
\unskip\
\newblock
\APACrefYearMonthDay{2010}{}{},
\newblock
\unskip
\newblock
\APACjournalVolNumPages{Class. Quantum Grav.}{27}{}{165003}.
\PrintBackRefs{\CurrentBib}

\bibitem [\protect \citeauthoryear {%
Friedman%
}{%
Friedman%
}{%
{\protect \APACyear {1922}}%
}]{%
Friedmann1922}
\APACinsertmetastar {%
Friedmann1922}%
\begin{APACrefauthors}%
Friedman, A.%
\end{APACrefauthors}%
\unskip\
\newblock
\APACrefYearMonthDay{1922}{}{},
\newblock
\unskip
\newblock
\APACjournalVolNumPages{Zeitschrift f\"ur Physik}{10}{}{377}.
\PrintBackRefs{\CurrentBib}

\bibitem [\protect \citeauthoryear {%
Hawking%
\ \BBA {} Hertog%
}{%
Hawking%
\ \BBA {} Hertog%
}{%
{\protect \APACyear {2018}}%
}]{%
Hawking2018}
\APACinsertmetastar {%
Hawking2018}%
\begin{APACrefauthors}%
Hawking, S.%
\BCBT {}\ \BBA {} Hertog, T.%
\end{APACrefauthors}%
\unskip\
\newblock
\APACrefYearMonthDay{2018}{}{},
\newblock
\unskip
\newblock
\APACjournalVolNumPages{High Energ. Phys.}{04}{}{147}.
\PrintBackRefs{\CurrentBib}

\bibitem [\protect \citeauthoryear {%
Hayward%
}{%
Hayward%
}{%
{\protect \APACyear {1994}}%
}]{%
Hayward}
\APACinsertmetastar {%
Hayward}%
\begin{APACrefauthors}%
Hayward, S\BPBI A.%
\end{APACrefauthors}%
\unskip\
\newblock
\APACrefYearMonthDay{1994}{}{},
\newblock
\unskip
\newblock
\APACjournalVolNumPages{Phys. Rev. D}{49}{}{6467}.
\PrintBackRefs{\CurrentBib}

\bibitem [\protect \citeauthoryear {%
Ijjas%
\ \BBA {} Steinhardt%
}{%
Ijjas%
\ \BBA {} Steinhardt%
}{%
{\protect \APACyear {2019}}%
}]{%
Ijjas2019}
\APACinsertmetastar {%
Ijjas2019}%
\begin{APACrefauthors}%
Ijjas, A.%
\BCBT {}\ \BBA {} Steinhardt, P.%
\end{APACrefauthors}%
\unskip\
\newblock
\APACrefYearMonthDay{2019}{}{},
\newblock
\unskip
\newblock
\APACjournalVolNumPages{Physics Letters B}{795}{}{666-672}.
\PrintBackRefs{\CurrentBib}

\bibitem [\protect \citeauthoryear {%
Ijjas%
, Steinhardt%
\BCBL {}\ \BBA {} Loeb%
}{%
Ijjas%
\ \protect \BOthers {.}}{%
{\protect \APACyear {2014}}%
}]{%
Ijjas2014}
\APACinsertmetastar {%
Ijjas2014}%
\begin{APACrefauthors}%
Ijjas, A.%
, Steinhardt, P.%
\BCBL {}\ \BBA {} Loeb, A.%
\end{APACrefauthors}%
\unskip\
\newblock
\APACrefYearMonthDay{2014}{}{},
\newblock
\unskip
\newblock
\APACjournalVolNumPages{Phys. Rev.}{D89}{}{023525}.
\PrintBackRefs{\CurrentBib}

\bibitem [\protect \citeauthoryear {%
Kornmesser%
}{%
Kornmesser%
}{%
{\protect \APACyear {2020}}%
}]{%
Kormesser2020}
\APACinsertmetastar {%
Kormesser2020}%
\begin{APACrefauthors}%
Kornmesser, M.%
\end{APACrefauthors}%
\unskip\
\newblock
\APACrefYearMonthDay{2020}{}{},
\newblock
\unskip
\newblock
\APACjournalVolNumPages{ESO}{}{}{},
\newblock
\APAChowpublished {Available at
  \url{https://supernova.eso.org/exhibition/1101/}}.
\PrintBackRefs{\CurrentBib}

\bibitem [\protect \citeauthoryear {%
Lema{\^i}tre%
}{%
Lema{\^i}tre%
}{%
{\protect \APACyear {1927}}%
}]{%
Lemaitre1927}
\APACinsertmetastar {%
Lemaitre1927}%
\begin{APACrefauthors}%
Lema{\^i}tre, G.%
\end{APACrefauthors}%
\unskip\
\newblock
\APACrefYearMonthDay{1927}{}{},
\newblock
\unskip
\newblock
\APACjournalVolNumPages{Annales de la Soci{\'e}t{\'e} Scientifique de
  Bruxelles}{A47}{}{49}.
\PrintBackRefs{\CurrentBib}

\bibitem [\protect \citeauthoryear {%
Mercuri%
}{%
Mercuri%
}{%
{\protect \APACyear {2010}}%
}]{%
Mercuri2010}
\APACinsertmetastar {%
Mercuri2010}%
\begin{APACrefauthors}%
Mercuri, S.%
\end{APACrefauthors}%
\unskip\
\newblock
\APACrefYearMonthDay{2010}{}{},
\newblock
\APACrefbtitle {Introduction to Loop Quantum Gravity.} {Introduction to Loop
  Quantum Gravity.}
\newblock
\APACrefnote{arXiv:1001.1330v1}
\PrintBackRefs{\CurrentBib}

\bibitem [\protect \citeauthoryear {%
Nielsen%
}{%
Nielsen%
}{%
{\protect \APACyear {2007}}%
}]{%
Nielsen}
\APACinsertmetastar {%
Nielsen}%
\begin{APACrefauthors}%
Nielsen, A\BPBI B.%
\end{APACrefauthors}%
\unskip\
\newblock
\APACrefYearMonthDay{2007}{10}{},
\newblock
\APACrefbtitle {Black holes as local horizons.} {Black holes as local
  horizons.}
\newblock
\APACrefnote{eprint arXiv:0711.0313}
\PrintBackRefs{\CurrentBib}

\bibitem [\protect \citeauthoryear {%
Nielsen%
\ \BBA {} Yoon%
}{%
Nielsen%
\ \BBA {} Yoon%
}{%
{\protect \APACyear {2008}}%
}]{%
Nielsen2008}
\APACinsertmetastar {%
Nielsen2008}%
\begin{APACrefauthors}%
Nielsen, A\BPBI B.%
\BCBT {}\ \BBA {} Yoon, J\BPBI H.%
\end{APACrefauthors}%
\unskip\
\newblock
\APACrefYearMonthDay{2008}{}{},
\newblock
\unskip
\newblock
\APACjournalVolNumPages{Classical and Quantum Gravity}{25 (8)}{}{085010}.
\PrintBackRefs{\CurrentBib}

\bibitem [\protect \citeauthoryear {%
Powell%
, Lopez%
\BCBL {}\ \BBA {} Matzner%
}{%
Powell%
\ \protect \BOthers {.}}{%
{\protect \APACyear {2020}}%
}]{%
Powell2020}
\APACinsertmetastar {%
Powell2020}%
\begin{APACrefauthors}%
Powell, J\BPBI R.%
, Lopez, R.%
\BCBL {}\ \BBA {} Matzner, R\BPBI A.%
\end{APACrefauthors}%
\unskip\
\newblock
\APACrefYearMonthDay{2020}{}{},
\newblock
\unskip
\newblock
\APACjournalVolNumPages{Entropy}{22 (7)}{}{795}.
\PrintBackRefs{\CurrentBib}

\bibitem [\protect \citeauthoryear {%
Robertson%
}{%
Robertson%
}{%
{\protect \APACyear {1935}}%
}]{%
Robertson1935}
\APACinsertmetastar {%
Robertson1935}%
\begin{APACrefauthors}%
Robertson, H.%
\end{APACrefauthors}%
\unskip\
\newblock
\APACrefYearMonthDay{1935}{}{},
\newblock
\unskip
\newblock
\APACjournalVolNumPages{Astrophysical Journal}{82}{}{248-301}.
\PrintBackRefs{\CurrentBib}

\bibitem [\protect \citeauthoryear {%
Vasconcellos%
, Hadjimichef%
, Hess%
, de Freitas~Pacheco%
\BCBL {}\ \BBA {} Bodmann%
}{%
Vasconcellos%
\ \protect \BOthers {.}}{%
{\protect \APACyear {2022}}%
}]{%
Zen2022}
\APACinsertmetastar {%
Zen2022}%
\begin{APACrefauthors}%
Vasconcellos, C\BPBI Z.%
, Hadjimichef, D.%
, Hess, P.%
, de Freitas~Pacheco, J.%
\BCBL {}\ \BBA {} Bodmann, B.%
\end{APACrefauthors}%
\unskip\
\newblock
\APACrefYearMonthDay{2022}{}{},
\newblock
\APACrefbtitle {Evidences for the Branch-Cut Cosmology.} {Evidences for the
  Branch-Cut Cosmology.}
\newblock
\APACrefnote{XXI Meeting of Physics. UNSAAC, Cusco, Per\'u, 16-18 December
  2021. To be published by Journal of Physics: Conference Series}
\PrintBackRefs{\CurrentBib}

\bibitem [\protect \citeauthoryear {%
Vasconcellos%
, Hadjimichef%
, Razeira%
, Volkmer%
\BCBL {}\ \BBA {} Bodmann%
}{%
Vasconcellos%
\ \protect \BOthers {.}}{%
{\protect \APACyear {2020}}%
}]{%
Zen2020}
\APACinsertmetastar {%
Zen2020}%
\begin{APACrefauthors}%
Vasconcellos, C\BPBI Z.%
, Hadjimichef, D.%
, Razeira, M.%
, Volkmer, G.%
\BCBL {}\ \BBA {} Bodmann, B.%
\end{APACrefauthors}%
\unskip\
\newblock
\APACrefYearMonthDay{2020}{}{},
\newblock
\unskip
\newblock
\APACjournalVolNumPages{Astronomische Nachrichten}{340 (9,10)}{}{857}.
\PrintBackRefs{\CurrentBib}

\bibitem [\protect \citeauthoryear {%
Vasconcellos%
\ \protect \BOthers {.}}{%
Vasconcellos%
\ \protect \BOthers {.}}{%
{\protect \APACyear {2021}}%
{\protect \APACexlab {{\protect \BCnt {1}}}}}]{%
Zen2021b}
\APACinsertmetastar {%
Zen2021b}%
\begin{APACrefauthors}%
Vasconcellos, C\BPBI Z.%
, Hess, P.%
, Hadjimichef, D.%
, Bodmann, B.%
, Razeira, M.%
\BCBL {}\ \BBA {} Volkmer, G.%
\end{APACrefauthors}%
\unskip\
\newblock
\APACrefYearMonthDay{2021{\protect \BCnt {1}}}{}{},
\newblock
\unskip
\newblock
\APACjournalVolNumPages{Astronomische Nachrichten}{342 (5)}{}{776-787}.
\PrintBackRefs{\CurrentBib}

\bibitem [\protect \citeauthoryear {%
Vasconcellos%
\ \protect \BOthers {.}}{%
Vasconcellos%
\ \protect \BOthers {.}}{%
{\protect \APACyear {2021}}%
{\protect \APACexlab {{\protect \BCnt {2}}}}}]{%
Zen2021a}
\APACinsertmetastar {%
Zen2021a}%
\begin{APACrefauthors}%
Vasconcellos, C\BPBI Z.%
, Hess, P\BPBI O.%
, Hadjimichef, D.%
, Bodmann, B.%
, Razeira, M.%
\BCBL {}\ \BBA {} Volkmer, G.%
\end{APACrefauthors}%
\unskip\
\newblock
\APACrefYearMonthDay{2021{\protect \BCnt {2}}}{}{},
\newblock
\unskip
\newblock
\APACjournalVolNumPages{Astronomische Nachrichten}{342 (5)}{}{765-775}.
\PrintBackRefs{\CurrentBib}

\bibitem [\protect \citeauthoryear {%
Walker%
}{%
Walker%
}{%
{\protect \APACyear {1937}}%
}]{%
Walker1937}
\APACinsertmetastar {%
Walker1937}%
\begin{APACrefauthors}%
Walker, A.%
\end{APACrefauthors}%
\unskip\
\newblock
\APACrefYearMonthDay{1937}{}{},
\newblock
\unskip
\newblock
\APACjournalVolNumPages{Proceedings of the London Mathematical
  Society}{42}{}{90}.
\PrintBackRefs{\CurrentBib}

\bibitem [\protect \citeauthoryear {%
Zyla%
\ \BBA {} et al.%
}{%
Zyla%
\ \BBA {} et al.%
}{%
{\protect \APACyear {2020}}%
}]{%
Zyla2020}
\APACinsertmetastar {%
Zyla2020}%
\begin{APACrefauthors}%
Zyla, P.%
\BCBT {}\ \BBA {} et al.%
\end{APACrefauthors}%
\unskip\
\newblock
\APACrefYearMonthDay{2020}{}{},
\newblock
\unskip
\newblock
\APACjournalVolNumPages{PTEP}{8}{}{083C01}.
\PrintBackRefs{\CurrentBib}

\end{thebibliography}

\end{document}